\documentclass[twoside,12pt]{article}
\usepackage{CJK}
\usepackage{indentfirst}
\usepackage{bm}
\usepackage{epsfig}

\usepackage[colorlinks,
            citecolor=blue,
            anchorcolor=green,
            menucolor=orange,
            linkcolor=red,
            filecolor=red,
            runcolor=pink,
            urlcolor=blue,
            frenchlinks=red]{hyperref}

\begin{document}

\begin{CJK*}{GBK}{song}

\thispagestyle{empty} \vspace*{0.8cm}\hbox
to\textwidth{\vbox{\hfill\huge\sf Commun. Theor. Phys.\hfill}}
\par\noindent\rule[3mm]{\textwidth}{0.2pt}\hspace*{-\textwidth}\noindent
\rule[2.5mm]{\textwidth}{0.2pt}


\begin{center}
\LARGE\bf Three-dimensional cytoplasmic calcium propagation with boundaries$^{*}$
\end{center}

\footnotetext{\hspace*{-.45cm}\footnotesize $^*$This project is partially  supported by the National Natural Science Foundation of China under Grant No. 11675228 and China postdoctoral Science Foundation  under Grant No. 2015M572662XB.}
\footnotetext{\hspace*{-.45cm}\footnotesize $^\dag$Corresponding author, E-mail: junhe@njnu.edu.cn }

\begin{center}
\rm Han-Yu Jiang$^{\rm 1,2)}$ \ and   \ Jun He$^{\rm 1)\dagger}$
\end{center}

\begin{center}
\begin{footnotesize} \sl
${}^{\rm 1)}$Department of Physics and Institute of Theoretical Physics, Nanjing Normal University,
Nanjing, 210097, China \\
${}^{\rm 2)}$Sino-U.S. Center for Grazingland Ecosystem Sustainability/Pratacultural Engineering Laboratory of Gansu Province/ Key Laboratory of Grassland Ecosystem,Ministry of Education/College of Pratacultural Science, Gansu Agricultural University, Lanzhou, 730070, CHina
\end{footnotesize}
\end{center}

\begin{center}
\footnotesize (Received XXXX; revised manuscript received XXXX)

\end{center}

\vspace*{2mm}

\begin{center}
\begin{minipage}{15.5cm}
\parindent 20pt\footnotesize
Ca$^{2+}$  plays  an important  role in cell  signal transduction. Its intracellular propagation  is the most basic process of  Ca$^{2+}$ signaling, such as calcium wave and double messenger system.  In  this work,  with both numerical simulation and mean field ansatz, the 3-dimensional probability distribution of Ca$^{2+}$, which is read out by phosphorylation, is studied in two scenarios with boundaries.   The coverage of distribution of Ca$^{2+}$ is found at an order of magnitude of $\mu$m, which is consistent with experimental observed calcium spike and wave.  Our results  suggest that the double messenger system may occur in the ER-PM junction to acquire great efficiency. The buffer effect of kinase is also discussed by calculating the average position of phosphorylations and free Ca$^{2+}$.  The results are helpful to understand the mechanism of  Ca$^{2+}$ signaling.
\end{minipage}
\end{center}

\begin{center}
\begin{minipage}{15.5cm}
\begin{minipage}[t]{2.3cm}{\bf Keywords:}\end{minipage}
\begin{minipage}[t]{13.1cm}
Ca$^{2+}$ signaling, Diffusion, Gillespie algorithm, Mean field ansatz
\end{minipage}\par\vglue8pt

\end{minipage}
\end{center}

\section{Introduction}

Cytoplasmic Ca$^{2+}$ is the most basic second messenger of cell signal transduction~\cite{Clapham2007,Berridge1998}. It  widely involves in the regulation of cell life activities, including respondence to external stimuli, cell membrane permeability, cell secretion, metabolism and differentiation~\cite{Berridge2016}. 
Within the cellular signaling network, the accurate decoding of diverse Ca$^{2+}$ signal is a fundamental molecular event~\cite{Berridge1983a,Streb1983b}. In the double messenger system,  the activity phospholipase C (PLC$\beta$)  induces the hydrolysis of the phospholipid phosphatidylinositol-4, 5-bisphosphate (PIP$_2$) to generate inositol 1,4,5-trisphosphate (IP$_3$) and diacylglycerol (DAG) in the plasma membrane (PM).  IP$_3$  mobilizes the endogenous Ca$^{2+}$, transfers the Ca$^{2+}$ stored in  endoplasmic reticulum  (ER) to the cytoplasm, and increases the intracellular Ca$^{2+}$ concentration. Ca$^{2+}$ combines with protein kinase C (PKC), which  is a kind of calcium dependent kinase, and participates in many physiological processes, short-term physiological effects such as cell secretion and muscle contraction, or long-term physiological effects such as cell proliferation and differentiation.  PKC  is translocated to the inner surface of the PM, which is activated by DAG, another second messenger of phosphatidylinositol signal  bound to PM.  Then, the serine and threonine residues of different substrate proteins in different cell types are phosphorylated.  In the current work, we define  the  response strength of  double messenger system to stimuli as the {\it efficiency} of such signal pathway, which is described by the phosphorylation events induced near the position of the stimuli in PM. 
Ca$^{2+}$ also combines with calmodulin (CaM)  to form a  Ca$^{2+}$-CaM complex, which  then  activates the target enzymes~\cite{Cheung1980}. CaM kinases are highly conserved in all eukaryotes, and are important target enzymes. It  mediates many functional activities in animal and plant cells, plays an important role in the synthesis of cAMP and cGMP and degradation of glycogen, and smooths muscle contraction and neurotransmitter secretion and synthesis. Furthermore, since  CaM may diffuse slowly with a speed smaller than Ca$^{2+}$, it plays a role of buffer of the Ca$^{2+}$ propagation to enlarge  its influence coverage~\cite{Palecek1999,John2001}.

The intracellular propagation and distribution of Ca$^{2+}$ ion affects the activation and operation of downstream components of signaling, such as  CaM and PKC~\cite{Yang2020}.  According to the amplitude of the Ca$^{2+}$ spikes, the frequency of the Ca$^{2+}$ waves, and the microdomains of the Ca$^{2+}$ flickers, Ca$^{2+}$ signaling displays different temporal and spatial patterns~\cite{Jaffe1993,Wei1992,Cheng2008,Dani1992}.   The distribution of intracellular Ca$^{2+}$ is strictly regionalized, while maintaining the homeostasis of cytoplasmic Ca$^{2+}$ is the prerequisite for normal cell growth~\cite{Bush1995}. Therefore, it is of great significance to  analyze the spatial distribution and dynamic changes of intracellular Ca$^{2+}$ for the study of cell physiological characteristics and signal transduction mechanism. 
In the past few decades, a variety of experimental methods were developed to measure  spatial distribution of Ca$^{2+}$ concentration  in cytoplasm, endoplasmic reticulum,  and other organelles. In Ref.~\cite{Yamashita1990}, using digital imaging microscopy and the dye fura-2,  distribution of intracellular cytoplasm-free Ca$^{2+}$ were studied. The laser scanning confocal microscope technology was also developed as a research tool for observing the spatial and temporal changes of intracellular calcium distribution~\cite{MOMalley1995,Oshima1996}.    Fluorescent indicators for Ca$^{2+}$ are important tool  to detect the Ca$^{2+}$~\cite{Miyawaki1997,Schaferling2012}. By using a laser scanning confocal microscope and the fluorescent calcium indicator fluo-3, Cheng and his collaborators detected the calcium sparks in  heart cells~\cite{Cheng1993,Cheng1996}. A plasmonic-based electrochemical impedance microscope  was also developed to provide valuable information without a  fluorescent labeling~\cite{Lu2015}.  On the other hand, the establishment of  spatial network model of calcium ion,  by biochemical reaction modeling or parameter fitting, is also important  to understand the spatial and temporal characteristics of calcium signal and the mechanism of signal transduction. In 2010, Rudiger and Shuai used deterministic and stochastic simulation methods to simulate a network model in which several IP$_3$ receptors  channels release local calcium signals in a single cluster~\cite{Rudiger2010}. Qi established a model determined by a simple calcium kinetic equation and a Markov process~\cite{Qi2014}.  The full width at half-maximum of calcium spikes was also analyzed  and well reproduced in the simulation~\cite{Tan2007}.  

Although a lot of work has been done to study  calcium spatial distribution, most of them are about  calcium concentration, which is important to understand the calcium spikes and calcium waves, which  usually propagate along the membrane. However, for the double messenger system, we need to know the propagation and diffusion of the calcium ion from EM to PM instead of concentration. Hence, the current understanding of the decoding mechanism of CaM  and the signal network of calcium PKC still is fragmentary.
Besides, little is known about the spatial location of above kinase phosphorylation events for calcium signal transduction study. In Refs.~\cite{Wasnik2019a,Wasnik2019b},  the authors developed a frame to estimate the accuracy of the position of  external stimuli. In their work, the Ca$^{2+}$ induced by stimuli enters the cytoplasm through a channel in membrane, and is read out by the  spatial distribution of phosphorylation events in cytoplasm or membrane. It is an important reaction mechanism of the external stimuli. However, the double messenger system happens between ER and PM, which is quite different from the scenario studied in Refs.~\cite{Wasnik2019a,Wasnik2019b}. Theoretically, the double messenger system should be a diffusion problem with two boundaries. Moreover, with the increase of the distance between ER and PM, the probability distribution of the Ca$^{2+}$ becomes small, which makes the signal propagation in such pathway more difficult.   The efficiency of the double messenger system should be dependent on the distance between ER and PM.  The study about such issue is scarce in the literature.  Hence, it is interesting to preform a study about the  propagation of the Ca$^{2+}$ perpendicular to the membrane through the phosphorylation. 

By studying the random binding of calcium ions with calmodulin and relevant kinases distributed in different cytoplasmic regions, we can estimate the position of phosphorylation events in the cytoplasm or in the PM, as well as the position of binding  of calcium ion to CaM / PKC and the influence domain of calcium ion through phosphorylation.
Due to the short-term characteristics of calcium puff and the complexity of intracellular environment, it is difficult and cumbersome to locate the distribution area and binding probability of calcium ions by biological experiments. It is interesting to introduce  theoretical model to simulate such process. Based on such new idea, by  adopting mathematical simulation and mean-filed ansatz to solve non-linear diffusion equation, we start from the level of single calcium ion to study: (1)  the spatial probability of Ca$^{2+}$  read out by  phosphorylation from ER to PM, which provides an overall picture of the  Ca$^{2+}$ distribution; (2) average positions of phosphorylation and the free Ca$^{2+}$ in a direction perpendicular to the membrane, which is related to  buffer effect of  CaM; (3)  estimation error of the readout position of phosphorylation event, which assesses the effective region of  Ca$^{2+}$.  Large efficiency of double messenger system benefits from large number of phosphorylation events  and small estimation error in PM. With these results, we can analyze the dependence of the efficiency of the double messenger system on the distance between ER and PM.  This study provides a theoretical basis for further exploring the mechanism of Ca$^{2+}$ mobilization induced calcium signaling pathway in cell biology, and the correlation between spatial heterogeneity of calcium ions and calcium signaling pathway.

\section{Theoretical frame}

\begin{figure}[h!]
  \centering
  \includegraphics[bb=30 0 680 520, clip,scale=0.5,angle=0]
  {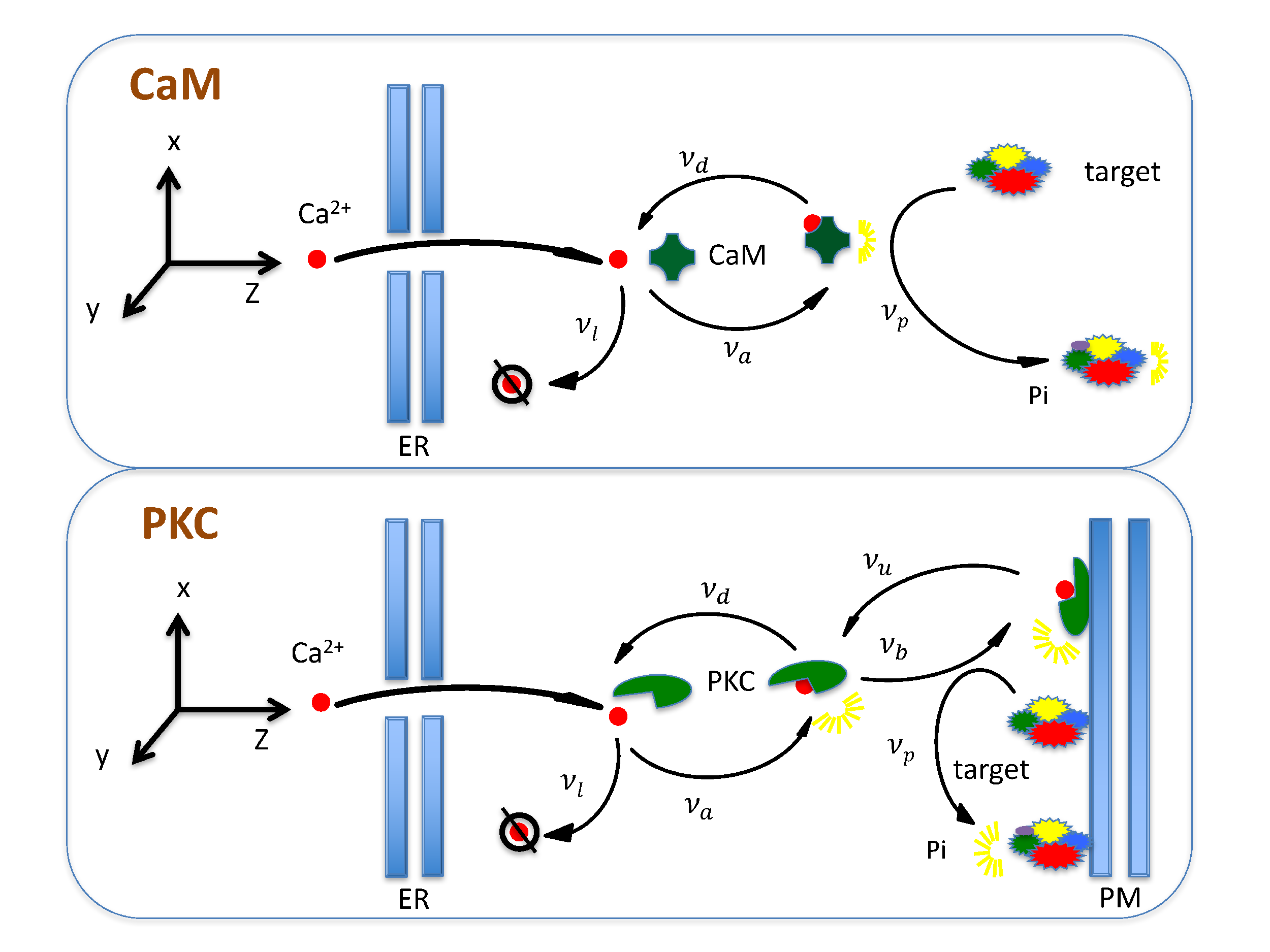}
  \caption{\small The CaM and PKC scenarios. The explicit description is presented in text.  }\label{FigCaM}
 \end{figure}

In the current work, we consider CaM and PKC scenarios  (see Fig.~\ref{FigCaM}).  Ca$^{2+}$ enters cytoplasm through calcium pump on  ER.  In  the CaM scenario, the phosphorylation happens in  cytoplasm  while  the PKC attached by Ca$^{2+}$ should be bound onto  PM first in PKC scenario. In the followings, we will describe these two scenarios explicitly when presenting   master equation and simulation method. 

\subsection{Master equation}

First, we present analytical description of Ca$^{2+}$  by the master equation, with which the spatial probability distribution of phosphorylation events is introduced. 

\subsubsection{Diffusion of free Ca$^{2+}$ until first attachment}

In the literature, the permeation of  Ca$^{2+}$  ion in water and cytosolic was discussed~\cite{Kuyucak2001}.  Moving  ion will be damped because its energy is lost to  surrounding medium through  viscous resistance to that movement. It can be described by using the Langevin equation~\cite{Chandrasekhar1943}, $m\frac{{\rm d}{\bm v}}{{\rm d}t}=-m\gamma{\bm v}+{\bm R}$
where the $m$, ${\bm v}$,  and  $\gamma$ are the mass, velocity and the friction coefficient of the ion, and ${\rm R}$ is  random force.  Average velocity deceases exponentially as $\langle{\bm v}\rangle=\langle{\bm v}\rangle_0{\rm e}^{-\gamma t}$. Based on the Einstein relation, the friction coefficient of the calcium ion can be evaluated as $\gamma^{-1}\approx 1$~fs~\cite{Hille1992}.  It suggests that  initial velocity of calcium ion will be lost in an extremely short time after entering  cytoplasm. Hence,  permeation of Ca$^{2+}$ behaves as a diffusion. 

We define a  probability distribution  function as ${\cal P}_{i_0}[{\bm r},t]$ for the state with $\rm Ca^{2+}$ ion unattached to a kinase, and we will call the Ca$^{2+}$ in such state as ``free Ca$^{2+}$" here and hereafter. It satisfies initial condition as
${\cal P}_{i_0}[{\bm r},t=0]=\delta({\bm r})$
which suggests that  $\rm Ca^{2+}$ ion enters  system at ${\bm r}=0$  on  ER in both scenarios, that is, we define  calcium pump as the origin of coordinate. 
If not losing into cytosolic,  the moving Ca$^{2+}$ ion attaches and activates a kinase, CaM or PKC 
, and it obeys  diffusion equation as, 
\begin{eqnarray}
\partial_t{\cal P}_{i_0}[{\bm r},t]=D_C\nabla^2{\cal P}_{i_0}[{\bm r},t]-(\nu_a+\nu_l){\cal P}_{i_0}[{\bm r},t],\label{i0}
\end{eqnarray}
where the $D_C$ is the diffusion constant for Ca$^{2+}$ in cytosolic, and $\nu_a$ and $\nu_l$ are the rates of attachment and losing, respectively.

 In Refs.~\cite{Wasnik2019a,Wasnik2019b},  Ca$^{2+}$ ion enters and  kinase binds to the same  PM, which is important in the determination of  position of external signal.  In the current work,  we consider 3-dimensional PKC scenario (see Fig.~\ref{FigCaM}), which happens in the important double messenger system. It leads to obvious difference.  Ca$^{2+}$ ion enters from  ER idealized as an infinite plane and diffuses in 3-dimensional space.  Furthermore, PKC needs to propagate to PM as another plane at a distance $l$. 
Hence, the $\rm Ca^{2+}$ ion will be reflected by boundary, PM and/or ER, which leads to boundary condition  as 
\begin{eqnarray}
\partial_z{\cal P}_{i_0}[{\bm r},t]|_{z=0}&=&0, \quad{\rm for \ CaM \  and \ PKC,}\label{bi01}\\
\partial_z{\cal P}_{i_0}[{\bm r},t]|_{z=l}&=&0,\quad{\rm for\ PKC\ only.}\label{bi02}
\end{eqnarray}

\subsubsection{Attachment and  detachment}

After the first attachment, a function ${\cal P}_{a}[n({\bm \xi}); {\bm r},t]$ is introduced to reflect two probabilities: having kinase activated by Ca$^{2+}$ ion  at position ${\bm r}$ and time $t$,  and having a distribution of phosphorylation event $n({\bm \xi})$ at time $t$.  ${\cal P}_{i}[n({\bm \xi}); {\bm r},t]$ is defined analogously but with  Ca$^{2+}$ detached from the kinase.
The master equation is given by
\begin{eqnarray}
\partial_t{\cal P}_i[(n({\bm \xi});{\bm r},t]=D_C\nabla^2{\cal P}_i[n({\bm \xi});{\bm r},t]+\nu_d{\cal P}_a[n({\bm \xi});{\bm r},t]-(\nu_a+\nu_l){\cal P}_i[n({\bm \xi});{\bm r},t],\label{i}
\end{eqnarray}
where $\nu_d$ is the rate of detachment.

 Because $\rm Ca^{2+}$ ion  and  active kinase will be also reflected by boundary, we have boundary condition  as 
\begin{eqnarray}
\partial_z{\cal P}_{i, a}[n({\bm \xi}); {\bm r},t]|_{z=0}&=&0, \quad{\rm for \ CaM \  and \ PKC,}\label{bi1}\\
\partial_z{\cal P}_{i}[n({\bm \xi}); {\bm r},t]|_{z=l}&=&0,\quad{\rm for\ PKC\ only.}\label{bi2}
\end{eqnarray}

\subsubsection{Phosphorylation in CaM scenario}

Difference appears after activation of kinase in two scenarios. In  CaM scenario (see Fig.~\ref{FigCaM}),  the active kinase phosphorylates directly in cytosolic at a rate of $\nu_p$, which can be described as
\begin{eqnarray}
\partial_t{\cal P}_a[n({\bm \xi});{\bm r},t]&=&D_K\nabla^2{\cal P}_a[n({\bm \xi});{\bm r},t]-\nu_d{\cal P}_a[n({\bm \xi});{\bm r},t]+\nu_a\left\{{\cal P}_i[n({\bm \xi});{\bm r},t]+{\cal P}_{i0}[{\bm r},t]\right\}\nonumber\\
&+&\nu_p\left\{{\cal P}_a[n({\bm \xi})-\delta({\bm \xi}-{\bm r});{\bm r},t]-{\cal P}_a[n({\bm \xi});{\bm r},t]\right\}.\label{aCaM}
\end{eqnarray}
The active kinase diffuses by constant $D_K$. The Dirac delta function $\delta({\bm \xi}-{\bm r})$ means that a phosphorylation event happens at ${\bm r}$,
then, the distribution jumps from ${\cal
P}_a[n({\bm \xi})-\delta({\bm \xi}-{\bm r});{\bm r},t]$ to ${\cal P}_a[n({\bm \xi});{\bm r},t]$ at a rate of $\nu_p$.  

\subsubsection{Binding and phosphorylation in PKC scenario}

In  PKC scenario,  phosphorylation does not happen directly after attachment. The PKC attached by a Ca$^{2+}$ diffuses and  obeys a master equation
\begin{eqnarray}
\partial_t{\cal P}_a[n({\bm \xi});{\bm r},t]=D_K\nabla^2{\cal P}_a[n({\bm \xi});{\bm r},t]-\nu_d{\cal P}_a[n({\bm \xi});{\bm r},t]+\nu_a\{{\cal P}_i[n({\bm \xi});{\bm r},t]+{\cal P}_{i_0}[{\bm r},t]\}.\label{aPKC}
\end{eqnarray}

The kinase should be bound onto  PM and activated  by DAG to make phosphorylation possible. A new state function ${\cal P}_b[n({\bm \xi});\bar{\bm r},t]$ is defined as 
the probability distribution, when $\rm Ca^{2+}$ attaches to a kinase and the complex binds into PM. Because PM is at a fixed distance from  ER, we adopt $\bar{\bm r}$ to denote the ${\bm r}$ with $z$ equaling to a distance $l$.  The binding and unbinding of PKC to PM can be written as a boundary condition is~\cite{Crank1975},
\begin{eqnarray}
-D_K\partial_z{\cal P}_a[n({\bm \xi});\bar{\bm r},t]=\nu_u{\cal P}_b[n({\bm \xi});\bar{\bm r},t]-\nu_b{\cal P}_a[n({\bm \xi});\bar{\bm r},t],\label{bu}
\end{eqnarray}
where $\nu_b$ and $\nu_u$ are the rates of binding and unbinding from PM, respectively. 

The phosphorylation at a rate of $\nu_p$ on  PM follows master equation as
\begin{eqnarray}
\partial_t{\cal P}_b[n({\bm \xi});\bar{\bm r},t]&=&\nu_b{\cal P}_a[n({\bm \xi});\bar{\bm r}, t]-\nu_u{\cal P}_b[n({\bm \xi});\bar{\bm r},t]\nonumber\\
&+&\nu_p\{{\cal P}_b[n(\bar{\bm \xi})-\delta(\bar{\bm \xi}-\bar{\bm r});\bar{\bm r},t]-{\cal P}_b[n(\bm \xi);\bar{\bm r},t]\}.\label{phPKC}
\end{eqnarray}

In both PKC and CaM scenarios, we define a probability distribution ${\cal P}_0[(n({\bm \xi});{\bm r},t]$ when the $\rm Ca^{2+}$ ion is lost, which obeys
\begin{eqnarray}
\partial_t{\cal P}_0[n({\bm \xi});t]=\nu_l\int {\rm d}{\bm r}\left\{{\cal P}_i[n({\bm \xi});{\bm r},t]+{\cal P}_{i_0}[{\bm r},t]\right\}.\label{0}
\end{eqnarray}

In summary, the CaM scenario shown in upper panel of Fig.~\ref{FigCaM} is described by diffusion equations in Eqs.~(\ref{i0},\ref{i},\ref{aCaM},\ref{0}) and boundary conditions in Eqs.~(\ref{bi01},\ref{bi1}), and the PKC scenario shown in lower panel of Fig.~\ref{FigCaM} is described by diffusion equations in Eqs.~(\ref{i0},\ref{i},\ref{aPKC},\ref{0}) and boundary conditions in Eqs.~(\ref{bi01},\ref{bi02},\ref{bi1},\ref{bi2},\ref{bu},\ref{phPKC}).
Hereafter, we will make a dimensionless treatment. The time and space are scaled by $\nu_p$ and $\sqrt{\nu_p/D_C}$, respectively. The scaled $\nu_a$, $\nu_d$, $\nu_l$, and $D_K$ can be obtained from scaled time and space.

\subsection{Stochastic simulation of  master equation}

With the master equation above, theoretically, we can obtain the distribution of phosphorylation event, $n(\bm \xi)$. However,  the master equations are difficult to be solved analytically.  In the current work, we adopt the Gillespie algorithm to do the stochastic simulation~\cite{Giuespie1977,Wasnik2019a,Wasnik2019b} in 3-dimensional space  with the boundary conditions.

The simulation follows scenarios shown in  Fig.~\ref{FigCaM}. Ca$^{2+}$ ion  enters  system at ${\bm r}=\bm 0$. An event occurs  in a step of time $\Delta t=-lnr/(\nu_a+\nu_l)$ with $r$ being a random number in a range from 0 to 1. This event is attachment or leaving system, which is determined by drawing a random number  from 0 to $\nu_l+\nu_a$. If the random number is smaller than $\nu_l$,  simulation  stops.    If not,  the position of  active kinase is changed in three directions by drawing three independent random numbers $\Delta {\bm r}$
from the Gaussian distribution with zero mean and variance $2\Delta t$.  If  obtained new position is out of the boundaries, it will be reflected by boundaries to a position in allowed region , $z>0$ for CaM scenario and $l>z>0$ for PKC scenario.   

In  CaM scenario,  kinase may phosphorylate target protein or detach in a new step of time. The type of event is still determined by drawing a random number. At the same time, the kinase moves randomly. If  phosphorylation happens, the position is recorded as ${\bm \xi}_i$. If the Ca$^{2+}$ ion detaches from the active kinase, simulation continues to a  new loop.    When  simulation ends, we have a trajectory of $n_0$ phosphorylation events. With $N$ simulations, we obtain $n$ trajectories with phosphorylation events.  We would like to remind that if a simulation does not raise any phosphorylation event, we do not count it into the number of trajectories $n$.

In PKC scenario,  after the kinase is activated, the situation is more complex because it has possibility to be bound onto  PM.  Let us consider an active kinase in $z_{old}$, which certainly locates between  two boundaries, goes to a new position $z_{new}$ after a random movement at time $\Delta t$. If the $z_{new}$ is still between the two boundaries, we draw a random number $r$ between 0 and 1. If $r<e^{-[(l-z_{new})(l- z_{old})]/(D_{K}\Delta t)}\nu_b\sqrt{\pi\Delta t}/(2\sqrt{D_K})$,  the
kinase is bound to the membrane~\cite{Erban2007,Andrews2004}. If the  $z_{new}$ is larger than $l$, the kinase is bound if $r>1-\nu_b\sqrt{\pi\Delta t}/(2\sqrt{D_K})$. Otherwise, it will be reflected.  If the $z_{new}$ is smaller than 0,  kinase is reflected only. If after reflecting, the kinase is still out of  boundaries, the above step should be repeated until it goes to a position between the boundaries.  After binding to the membrane, a random number is drawn to 
determine whether phosphorylation happens or  kinase unbinds from the membrane. The position of phosphorylation will be recorded as in CaM scenario.

\subsection{Mean-field ansatz}

Now we consider mean-field ansatz (MF)
where  phosphorylation rate  at position ${\bm r}$ is assumed to be proportional
to the probability of finding a kinase at this position.
In CaM scenario, the expected number $\hat{n}({\bm r})$ of phosphorylation events  in the limit $t\to\infty$ is $\hat{n}({\bm r})=\bar{p}_a({\bm r})=\int^\infty_0 dt'p_a({\bm r},t')$, where the barred quantities indicate  time-integrated quantities~\cite{Wasnik2019a,Wasnik2019b} . 
For Eqs.~(\ref{i0}, \ref{i}, \ref{aCaM}), to obtain the distribution of  expected phosphorylation events, we are then left with solving, 
\begin{eqnarray}
  -\delta({\bm r})&=&\nabla^2\bar{p}_{i_0}({\bm r})-(\nu_a+\nu_l)\bar{p}_{i_0}({\bm r}),\nonumber\\
  0&=&\nabla^2\bar{p}_{i}({\bm r})+\nu_d\bar{p}_a({\bm r})-(\nu_a+\nu_l)\bar{p}_i({\bm r}),\nonumber\\
0&=&D_K\nabla^2\bar{p}_a({\bm r})-\nu_d\bar{p}_i({\bm r})+\nu_a[\bar{P}_{i}({\bm r})+\bar{P}_{i0}({\bm r})].\label{CAMmean}
\end{eqnarray}
The first equation can be easily solved as
\begin{eqnarray}
 \bar{p}_{i_0}(r) =e^{-\omega r}/2\pi r\equiv\hat{n}_0({\bm r}),\label{free}
\end{eqnarray}
which is also the probability distribution of last position of free Ca$^{2+}$. 

Now, we need solve the two equations left which are coupled to each other. The $z$ is larger than zero for these two equations. However, due to the boundary at $z=0$ is reflective, it is reasonable to make an even extension. Hence, we can preform the Fourier transformation for three direction as,
\begin{eqnarray}
0&=&-D_K{q}^2 \bar{p}_{a}({\bm q})-\nu_d\bar{p}_{a}({\bm q})+\nu_a[\bar{p}_{i}({\bm q})+\bar{p}_{i_0}({\bm q})],\nonumber\\
0&=&-{q}^2\bar{p}_{i}({\bm q})+\nu_d\bar{p}_{a}({\bm q})-\omega^2\bar{p}_{i}({\bm q}).
\end{eqnarray}
After inverse transformation, we have
\begin{eqnarray}
\hat{n}({\bm r})=\bar{p}_{a}(r)=\frac{\nu_a}{D_K}\frac{1}{\lambda_+^2-\lambda_-^2}\frac{e^{-\lambda_-r}-e^{-\lambda_+r}}{2\pi r},\label{distrCaM}
\end{eqnarray}
where $\lambda_\pm=\sqrt{(a\pm\sqrt{a^2-4b})/2}$ with $a=(\nu_a+\nu_l)+{\nu_d}/{D_K}$ and $b={\nu_l\nu_d}/{D_K}$.

As in the CaM scenario, the distribution of phosphorylation events in PKC scenario at time $t\to\infty$ is given by $\hat{n}=\bar{p}_b(\bar{\bm r})$, which is related to  kinase distribution as,
$\partial_t p_b(\bar{\bm r})=\nu_bp_a(\bar{\bm r})-\nu_up_b(\bar{\bm r})$ from Eq.~(\ref{bu}).
After integrating, we have 
\begin{eqnarray}
\hat{n}_{PKC}=\bar{p}_b(\bar{\bm r})=\frac{\nu_b}{\nu_u}\bar{p}_a(\bar{\bm r}).\label{PKCpk}
\end{eqnarray}

One can finds that MF master equation in PKC scenario  are the same as Eq.~(\ref{CAMmean}) in  CaM scenario. Difference is that PKC scenario has additional boundary condition at $z=l$. It makes the Fourier transformation adopted in  CaM scenario failure to solve these coupled differential equations because it only works in a region from 0 to $\infty$. Hence, in this work, we do not try to give analytical results, but present the simulation results in PKC scenario.   However, for large $l$,  effect of  boundary at $z=l$ on  master equations becomes small. If neglecting the boundary effect, the $p_a(\bar{\bm r})$ can be solved as in the CaM scenario. Combined with Eq.~(\ref{PKCpk}), one can expect a similar result except a factor for PKC scenario at large $l$. It can be used to check our simulation results.

\section{Results}
With above preparation,  spatial probability distribution of phosphorylation can be obtained, which will be presented first. Based on such distribution, we will analyze average position and estimation error also. 

\subsection{Spatial probability distribution of phosphorylation}
\begin{figure}[h!]
  \centering
  \includegraphics[scale=0.395]{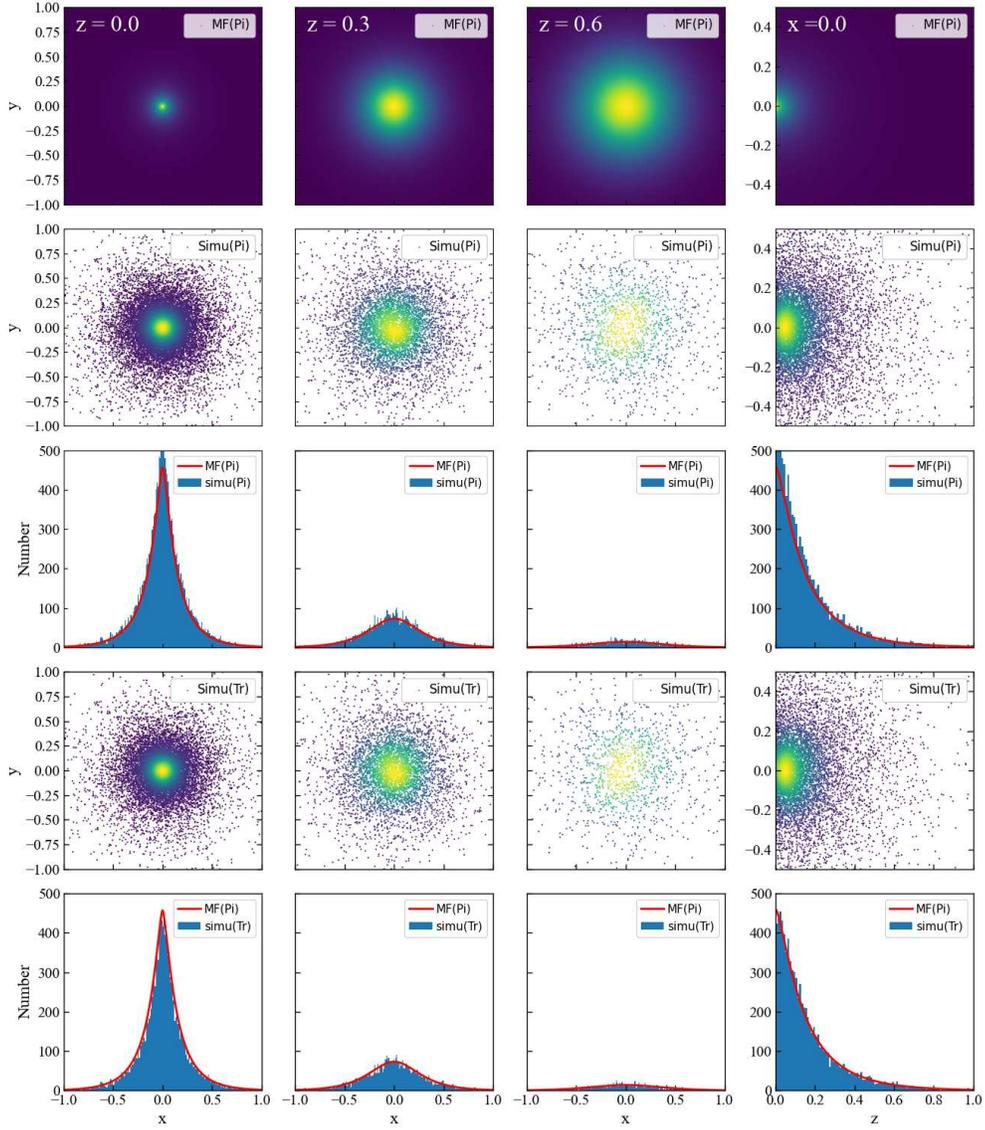}
  \caption{\small Spatial probability distribution of phosphorylation with $\nu_a$=10, $\nu_d$=10, $\nu_l$=20, and $D_K$=0.02~\cite{Nalefski2001,Smith1998,Donahue1987,Schaefer2001} from $10^7$ stochastic simulations in  CaM scenario. The panels in the first row are for results  of phosphorylation events (Pi) in the mean field ansatz (MF). The panels in the second and fourth rows are for the simulation (Simu) results of phosphorylation events (Pi) and trajectories (Tr), respectively. And comparisons between simulation and mean field ansatz are given in third and fifth rows. The color of the scattering points is determined by Gaussian kernel density estimation. The first three columns are for  slices at $z=0.0$, 0.3, and 0.6, with thickness $\delta z=0.01$,  and the last column is for $x=0$ with $\delta x=0.01$.  }\label{CaM0}
 \end{figure}

\begin{figure}[h!]
  \centering
  \includegraphics[scale=0.395]{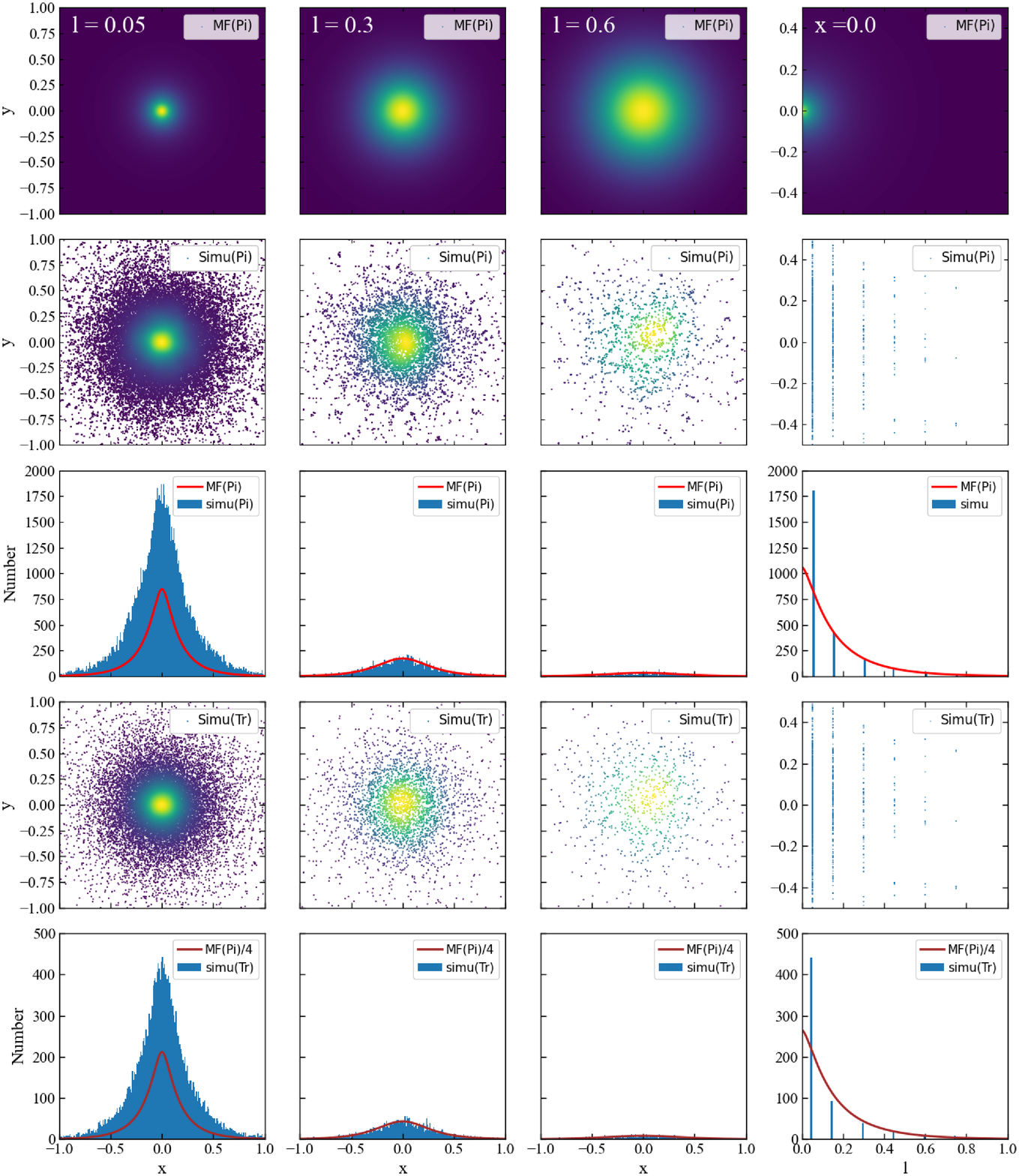}
  \caption{\small Spatial probability distribution of phosphorylation  with $\nu_a$=10, $\nu_d$=10, $\nu_l$=20, $\nu_b=\nu_u=1$, and $D_K$=0.02~\cite{Nalefski2001,Smith1998,Donahue1987,Schaefer2001} from $2\times10^5$ stochastic simulations in PKC scenario. The first three columns are for distances at $l=0.05$, 0.3, and 0.6, with thickness $\delta z=0.01$,  and the last column is for $x=0$ with $\delta x=0.01$. Other notations are analogous to those in Fig.~\ref{CaM0}.}\label{PKC}
 \end{figure}

Because spatial probability distribution of phosphorylation is in 3-dimensional space, we will present the results with slice map.  In Figs.~\ref{CaM0} and \ref{PKC}, we present the results for CaM scenario. Here, the values of parameters are cited from Refs.~\cite{Nalefski2001,Smith1998,Donahue1987,Schaefer2001}, which were measured in experiment.

In the first row of Fig.~\ref{CaM0}, the distribution with mean field  (MF) ansatz in Eq.~(\ref{distrCaM}) is presented. At small $z$, the phosphorylation concentrates at entrance. When deviating from ER,  phosphorylation events distribute in a larger region.  In the second and fourth rows, we present  simulation results for both phosphorylation events and trajectories.   In order to determine  event distribution at $z$, we  select events (Pi) at position ${\bm\xi}$ or trajectories (Tr) at mean position $\hat{\bm\xi}$ which are in a range form   $z-\delta z/2$ to $z+\delta z/2$. The simulation results exhibit the same picture as the MF ansatz.  To give a more obvious comparison, we present the results accumulated at $x$ direction in the third and fifth rows. The MF results are given by $\hat{n}({\bm r})\delta z\delta x N$, which is for events not trajectories.
The results suggest that the MF results for events fit the simulation as expected. The phosphorylation events concentrates at the entrance, and diminishes rapidly with the increase of $z$ and $x$.  It is interesting to observe that the results for  trajectories are almost the same as these for events. It suggests that  with the current parameters adopted, the number of the events in a trajectory is very small,  which ensures the effectiveness of MF ansatz. 
   
 Different from  CaM scenario,  all phosphorylations happen on PM in PKC scenario. In Fig~\ref{PKC}, we present the results with different distances,  $l=0.05$, $0.3$ and 0.6 between PM and ER. The simulations can not performed when PM and ER overlaps. Here we adopt a small distance $l=0.05$ instead of zero. 
At small distance, phosphorylations concentrate at center as in CaM scenario. With the increase of distance, the numbers of events and trajectories  diminish rapidly. As shown in third and fifth rows, the  phosphorylation events and trajectories are relatively more than in CaM scenario. As shown in the last column, the number is about two times large than the line for CaM, that is, $\hat{n}({\bm r})\delta x N$. At larger distance, the MF lines in CaM scenario fit the PKC simulation very well as expected.  Different from the CaM scenario, the number of trajectories is about four times larger than the number of events. In the last rows, we give the scaled results for MF, it is interesting to see that the lines also fit the simulation results very well at large distances. It suggests that the MF ansatz is still established, though a trajectory contains about 4 events averagely.

\subsection{Average position of  phosphorylation and  buffer effect of  kinase}

\begin{figure}[h!]
  \centering
  \includegraphics[scale=2.5]{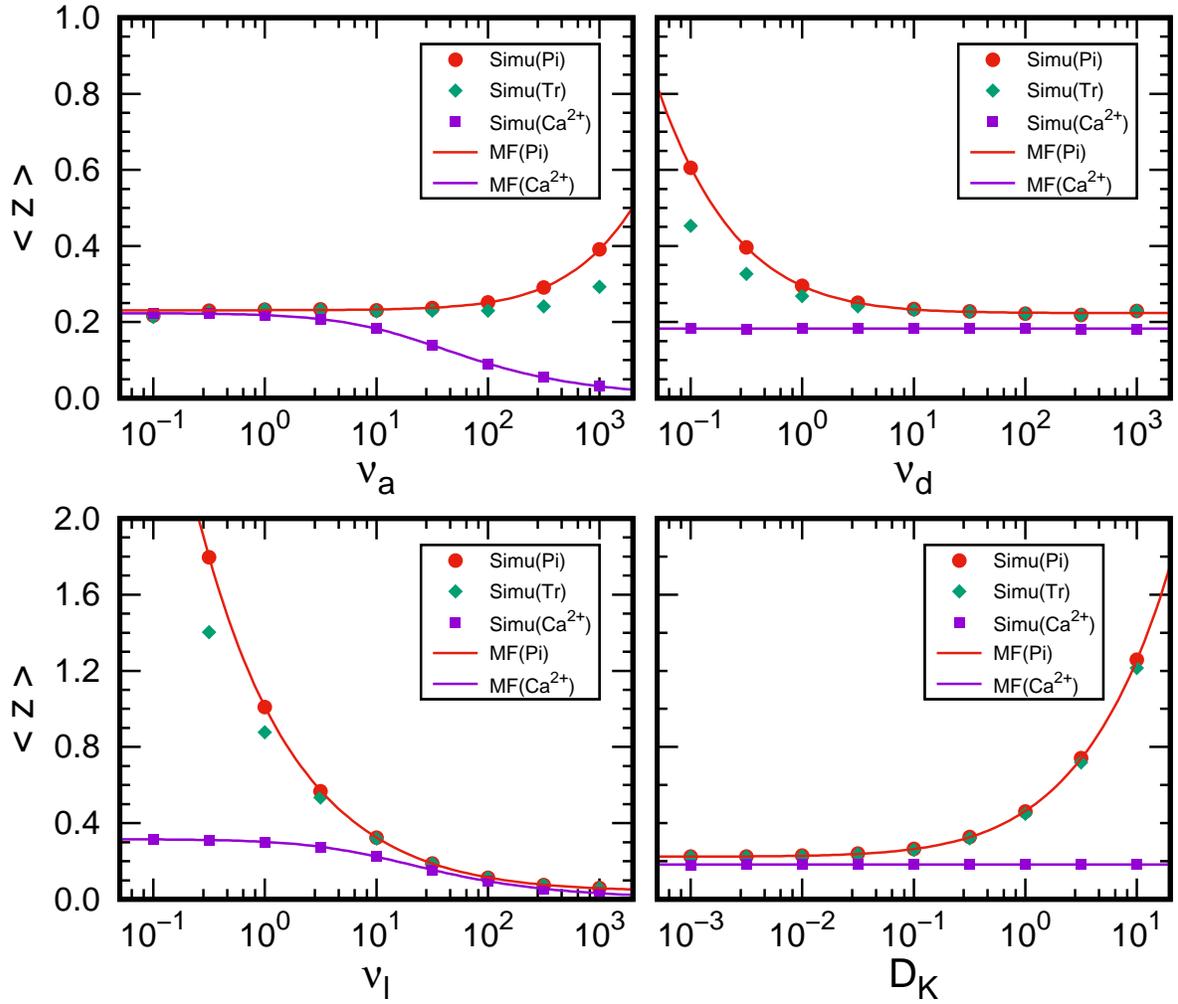}
  \caption{\small Average positions of  phosphorylation and  free Ca$^{2+}$  in CaM scenario.   The results are obtained by  varying one of the parameters,  $\nu_a$=10, $\nu_d$=10, $\nu_l$=20, and $D_K$=0.02.}\label{CaMline}
 \end{figure}

As shown in the above, phosphorylation consternates near  ER in CaM scenario, and the  number of phosphorylation events in PKC scenario is also much larger at small distance than  at large distance.  To give a numerical description, here we present  average position of phosphorylation in CaM scenario in Fig.~\ref{CaMline}. Because  average positions in $x$ and $y$ directions are zero, we only give $\langle z\rangle$ here. In  mean field ansatz,  it is defined as 
\begin{eqnarray}
\langle z\rangle=\frac{\int_{-\infty}^\infty dxdy\int^\infty_0dz~z~\hat{n}({\bm r})}{\int_{-\infty}^\infty dxdy\int^\infty_0dz~\hat{n}({\bm r})}=\frac{\lambda_-^2+\lambda_-\lambda_++\lambda_+^2}{\lambda_-^2\lambda_++\lambda_+\lambda_+^2}.
\end{eqnarray}
In simulation, we will collect all positions of events or mean positions of trajectories  $\xi_{zk}$ and the average position $\langle z\rangle=\sum_{k=1}^m \xi_{zk}/m$  with $m$ is the total number of events or trajectories.
 
With the parameters adopted in previous subsection, the average $\langle z\rangle$ is about  0.2, which corresponds to about 3 $\mu$m before dimensionless treatment. Such value is consistent with the order of the magnitude of a calcium spark~\cite{Cheng2008}. We also discuss the dependence of  $\langle z\rangle$ on the parameters.  The values are very stable with a rate of attachment  $\nu_a<100$,  a rate of detachment $\nu_d>1$, and a diffusion constant $D_K<0.1$. The result is more sensitive to the rate of loosing $\nu_l$.  With larger $\nu_a$ and smaller $\nu_d$, the Ca$^{2+}$ will attached into the CaM more quickly and enduringly, which reflects by larger $\langle z\rangle$. The large $D_K$ of the kinase also leads to larger   $\langle z\rangle$.
Such results suggest that the CaM is  buffer of Ca$^{2+}$ diffusion as suggested in Refs.~\cite{Palecek1999,John2001}. 

To give the buffer effect of CaM more clearly, we also present the average position of free Ca$^{2+}$ when it attaches to kinase or leaving the system before an attachment as described in Eq.~(\ref{free}). The average position with MF ansatz can be obtained as 
\begin{eqnarray}
\langle z\rangle=\frac{\int_{-\infty}^\infty dxdy\int^\infty_0dz~z~\hat{n}_0({\bm r})}{\int_{-\infty}^\infty dxdy\int^\infty_0dz~\hat{n}_0({\bm r})}=\frac{1}{\omega}.\label{freez}
\end{eqnarray}
The simulation results are also presented and compared. We would like to note that the $\langle z\rangle$ for free Ca$^{2+}$ is only dependent on $\nu_a$ and $\nu_l$.  In the stable region, the buffer effect is relatively small. When the $\langle z\rangle$ increases, the buffer effect of the CaM becomes important, and the phosphorylations will happen at a position far away from the ER, for example, with $\nu_l=1$, the average position of phosphorylations is at about 30 $\mu$m while the $\langle z\rangle$ for free Ca$^{2+}$ is only about 5 $\mu$m.   Larger $\nu_a$ and $D_K$, small $\nu_d$ and $\nu_l$ are beneficial to the buffer function of CaM.

\begin{figure}[h!]
  \centering
  \includegraphics[scale=2.6]{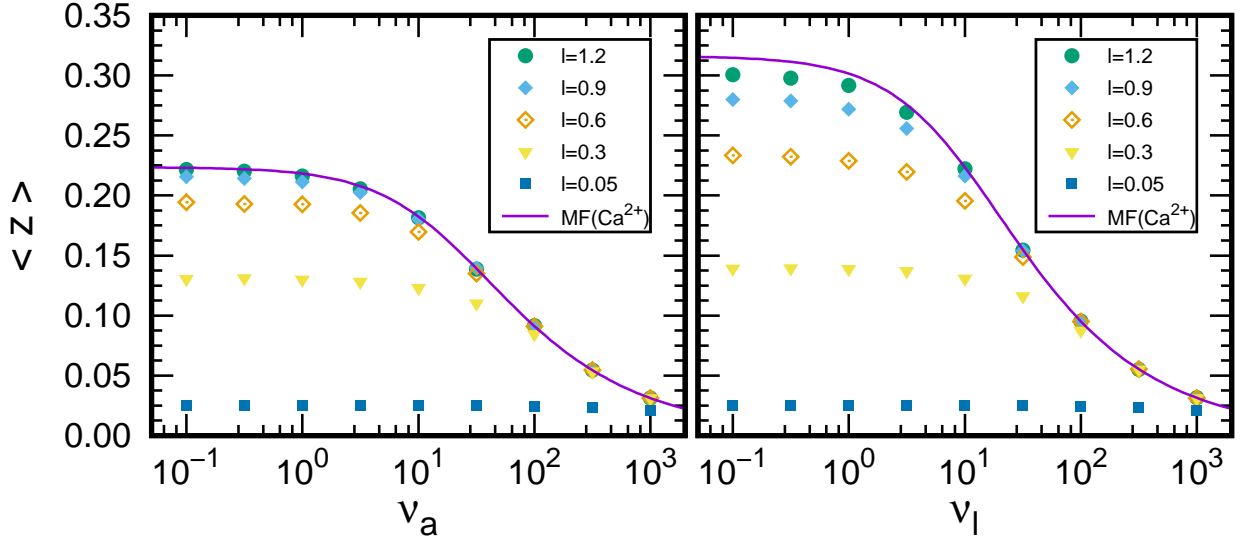}
  \caption{\small Average position of  free Ca$^{2+}$ in PKC scenario.   The results are obtained by  varying one of the parameters,  $\nu_a$=10, $\nu_d$=10, $\nu_l$=20, and $D_K$=0.02.}\label{PKCline}
 \end{figure}

In PKC scenario, the average position of phosphorylation is fixed at the PM. Here, we only give the results for  free Ca$^{2+}$ and compare them with CaM results with MF ansatz as shown in Fig.~\ref{PKCline}.
Since Ca$^{2+}$ will be reflected by PM and ER, the distance between two membranes will effect the average position for free Ca$^{2+}$. For a large  distance $l=1.2$, as expected the effect of  PM becomes small, the results are almost the same as the results in CaM scenario with MF ansatz.   In such cases, the buffer effect of the PKC is obvious.  Phosphorylations happen at PM which distance is about six time larger than the $\langle z\rangle$ for free Ca$^{2+}$ with $\nu_a<1$. With decrease of the distance, the $\langle z\rangle$ becomes small due to the suppression of  PM. Meantime, the buffer effect of the kinase becomes relatively  small too. For example, at a distance $l=0.3$, the $\langle z\rangle$ is about 0.15 with $\nu_a<1$.  With increase of both $\nu_a$ and $\nu_l$, Ca$^{2+}$ will be attached or lost quickly, so the buffer effect becomes more important.

\subsection{Estimation error of readout position of Ca$^{2+}$}

In the above, we present the average position in $z$ direction. In $x$ and $z$ directions,  average position is zero. Here we will give the estimation error to describe the distribution of phosphorylation parallel to ER.  With the mean field ansatz it is written as 
\begin{eqnarray}
 \ell^2&=&\frac{\int_{-\infty}^\infty dxdy~(x^2+y^2)~\hat{n}({\bm r})}{\int_{-\infty}^\infty dxdy~\hat{n}({\bm r})}\nonumber\\
&=&\frac{2e^{\lambda_-z}\lambda_-^3(1+\lambda_+z)-2e^{\lambda_+z}\lambda_+^3(1+\lambda_-z)}{\lambda_-^2\lambda_+^2(e^{\lambda_- z}\lambda_--e^{\lambda_+z}\lambda_+)}.\label{CaMell}
\end{eqnarray}
In simulation,  in order to determine  estimation error at $z$, we will collect the mean position of a simulation $\hat{\bm\xi}$ which satifies $z-\delta z/2<\hat{\xi}_{z}<z+\delta z/2$, with a number $n_z$. The number distribution is defined as $\hat{n}(z)={n_z}/{(n\delta z)}$ and the estimation error as $\ell^2(z)=\sum_{k=1}^{n_z} (\hat{\xi}^2_{x,k}+\hat{\xi}^2_{y,k})/n_z$. Here, the number distribution  $\hat{n}(z)$ is for the phosphorylation trajectories, not for the phosphorylation events. 

\begin{figure}[h!]
  \centering
  \includegraphics[scale=2.5]{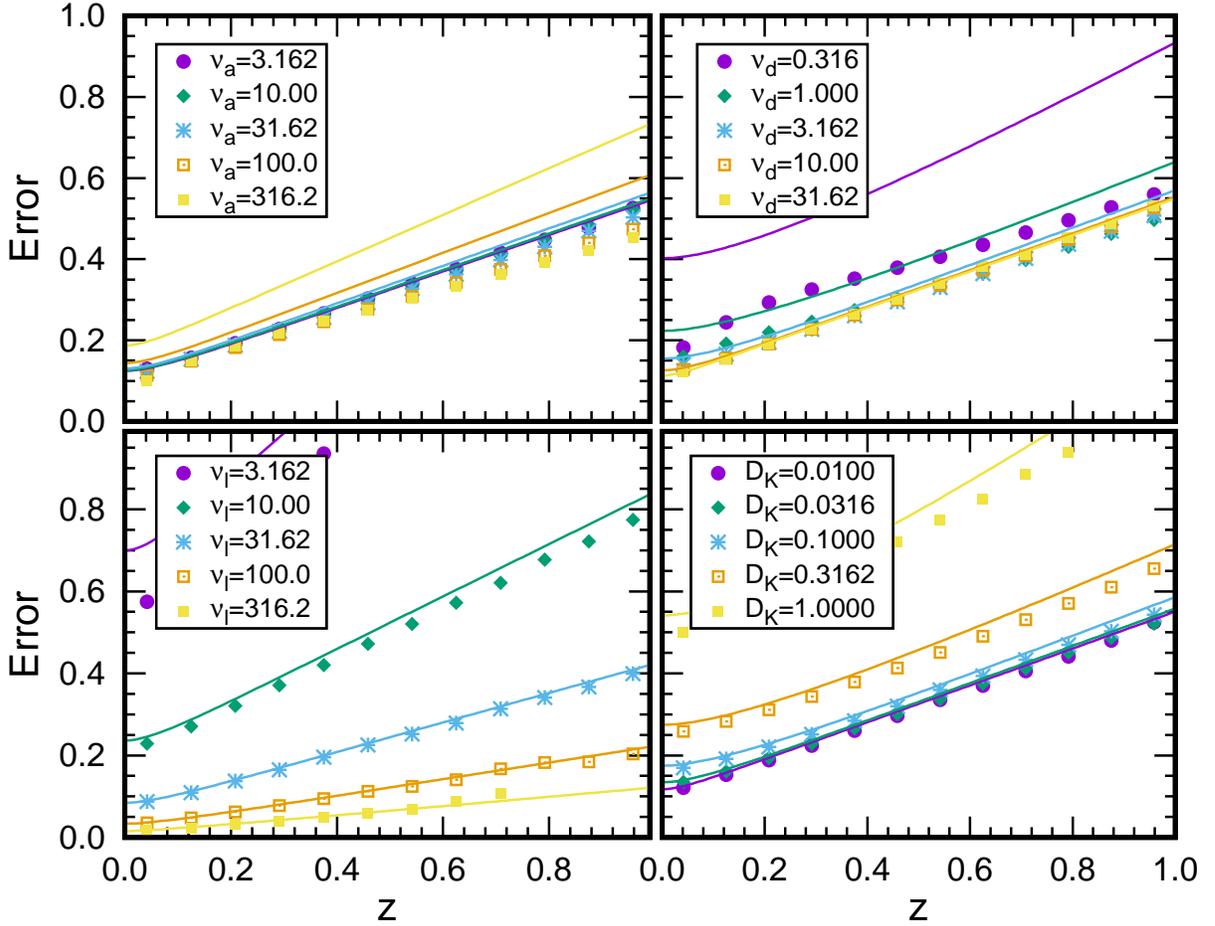}
  \caption{\small The variation of  estimation error against  distance with different parameters from $10^8$ stochastic simulations in CaM scenario.  The results are obtained by  varying one of the parameters,  $\nu_a$=10, $\nu_d$=10, $\nu_l$=20, and $D_K$=0.02. }\label{CaMerr}
 \end{figure}

We present CaM results for the estimation error in Fig.~\ref{CaMerr}. Dependence of  parameters and  comparison between  MF and simulation results are also presented.
For all cases,  estimation error increases with the increasing of $z$. The results are not sensitive to  variation of  $\nu_a$ in a large region from 3 to 300. For $\nu_a$ larger than 100, the simulation results begin to deviate from  MF results. Dependence on the  parameter  $\nu_d$ is also small, and appears at small value of   $\nu_d$.  The effect of $\nu_l$ is quite large from 3 to 300 while the MF and simulation results fit each other when $\nu_l$ larger than 10. 
The estimation error becomes larger with the increasing of diffusion parameter $D_K$.  Estimation error $\ell^2$ is about 0.2 at the $z=0.2$ with parameters we adopted, which corresponding 5 $\mu$m.

\begin{figure}[h!]
  \centering
  \includegraphics[scale=2.5]{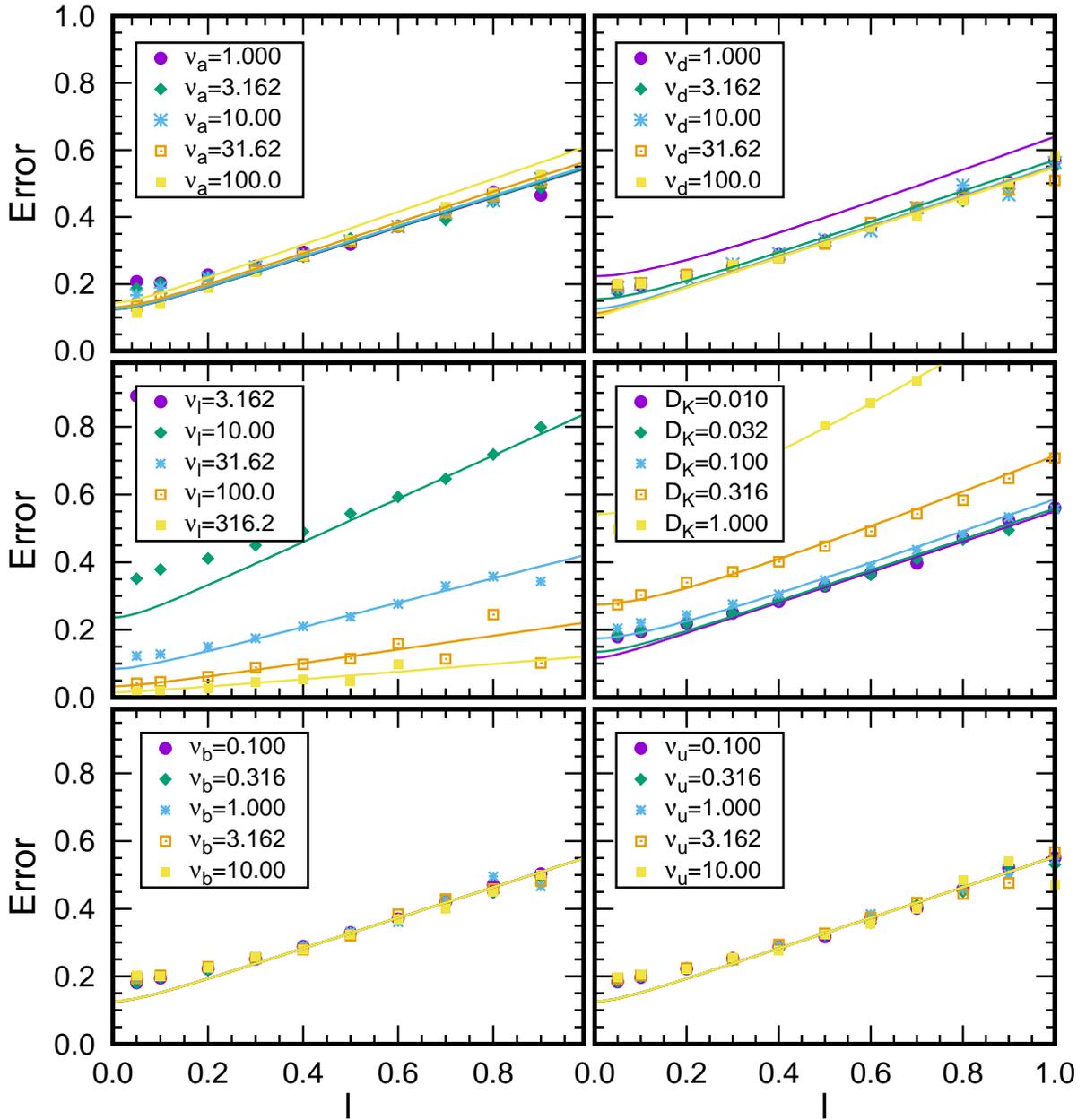}
  \caption{\small The variation of estimation error against  distance with different parameters from $10^8$ stochastic simulations in the PKC scenario.  Except the varied parameter, $\nu_a$=10, $\nu_d$=10, $\nu_l$=20, $\nu_b=1$, $\nu_u=1$, and $D_K$=0.02. }\label{PKCerr}
 \end{figure}

In Fig.~\ref{PKCerr}, We present PKC results for the estimation error, and compare it with the analytical results with Eq.~(\ref{CaMell}) in the CaM scenario. The results suggest the PKC simulation fit MF results in  CaM scenario at distance $l>0.2$ as expected. Generally speaking, the PKC results exhibit the same picture as the CaM scenario. Besides,  estimation error is independent on $\nu_b$ and $\nu_u$ as expected.

\section{Summary and discussion}

Ca$^{2+}$ signal is very important in cell signal transduction, which attracts much attentions from  scientific communities in biology, medicine, and physics. For all the research topics of the Ca$^{2+}$ signal, such as calcium spike, calcium wave, and second messenger system, the propagation of Ca$^{2+}$ ion in cytoplasm and its readout are in the most basic level. The study of spatial probability distribution of a Ca$^{2+}$ ion is helpful to understand intracellular  Ca$^{2+}$ signal transduction. In the current work, we study three-dimensional  Ca$^{2+}$ propagation in two typical CaM and PKC scenarios, and the phosphorylation is adopted to read out  Ca$^{2+}$. 

In CaM scenario, after released from the calcium channel in ER,  Ca$^{2+}$  diffuses in  cytoplasm, which corresponds to realistic case, such as a Ca$^{2+}$ event in calcium spike and wave~\cite{Cheng2008}.  In such case, Ca$^{2+}$ takes effect near membrane,  and effect of other membranes is very small. The result suggests that the probability distribution is only dependent of the radius from entrance.  With the parameters chosen~\cite{Nalefski2001,Smith1998,Donahue1987,Schaefer2001}, the phosphorylation events concentrate at the entrance and diminish exponentially  with the radius.  At the direction perpendicular to the membrane, the average position is only about several $\mu$m, which is at the same order of magnitude of the width of the calcium spike. The CaM is considered as buffer of the Ca$^{2+}$ propagation in cytoplasm~\cite{Palecek1999,John2001}.  With the parameters chosen, the buffer effect is not very large. However, there exist many types of CaM which has different properties. The results suggests that the average position in $z$ direction will increase with the increase of the rate of attachment $\nu_a$ and diffusion constant $D_K$, and decrease of the rate of detachment $\nu_d$ and the rate of losing $\nu_l$. At the direction parallel to  membrane, the average position is zero,  error  estimation $\ell^2$ is calculated to show the coverage of influence of a Ca$^{2+}$ ion.  At small $z$, the error  estimation is also at an order of several $\mu$m, and it will increase with the increase of $z$ almost linearly.  Generally speaking, if Ca$^{2+}$ is read out by phosphorylation,  coverage of influence of a Ca$^{2+}$ ion is at an order of $\mu$m.  The rates and diffusion constant of the CaM will effect  coverage of influence of Ca$^{2+}$ as a buffer. We would like to note that the results in CaM scenario can be applied to all type of membranes, not only the ER considered here.
   
The PKC scenario is quite different from the CaM scenario. In such scenario, the PKC attached by Ca$^{2+}$ must bind into the PM, which is the important part of the double messenger system.  All phosphorylation events happen in PM, and concentrate at zero point.  When the distance between ER and PM becomes smaller, the  phosphorylation events will decrease very rapidly, especially at small distance. At the same time, the estimation error  becomes smaller also. If we recall that the PKC should bind into the PM and activates by the DAG, a small distance  between PM and ER, smaller than 1~$\mu$m suggested by current result,  will benefit the  double message system.  In fact, the importance of the 
ER-PM junction were studied in many works, such as Ca$^{2+}$  influx~\cite{Jing2015,Prakriya2015}. The current work suggest that in the ER-PM junction, the efficiency of the double messenger system may be improved greatly.
   
In  summary, the 3-dimensional probability distribution of Ca$^{2+}$ which is read out by phosphorylation in cytoplasm is studied in two scenarios.  The results in PKC scenario with two boundaries are found consistent with the results in CaM scenario especially at large distance as suggested by MF ansatz.  The coverage of distribution of Ca$^{2+}$ is found at an order of $\mu$m, which is consistent with experimental observed calcium spike and wave.  Our results also suggest that the double messenger system may occurs in the ER-PM junction to acquire great efficiency.

\end{CJK*}
\end{document}